\documentstyle[12pt,epsfig]{article}
\begin{document}
\setlength{\baselineskip}{0.30in}
\newcommand{\beq}{\begin{equation}}
\newcommand{\eeq}{\end{equation}}
\newcommand{\bi}{\bibitem}

{\hbox to\hsize{March, 1997 \hfill TAC-1997-10}
\begin{center}
\vglue .06in
{\Large \bf {Nonequilibrium Corrections to the Spectra of Massless
 Neutrinos in the Early Universe. }
}
\bigskip
\\{\bf A.D. Dolgov
\footnote{Also: ITEP, Bol. Cheremushkinskaya 25, Moscow 113259, Russia.}
\footnote{e-mail: dolgov@tac.dk}, 
S.H. Hansen\footnote{e-mail: sthansen@tac.dk}
 \\[.05in]
{\it{Teoretisk Astrofysik Center\\
 Juliane Maries Vej 30, DK-2100, Copenhagen, Denmark
}}}
\\{\bf D.V. Semikoz\footnote{e-mail: semikoz@ms2.inr.ac.ru}}\\
{\it{Institute of Nuclear Research of the Russian Academy of Sciences\\
 60th October Anniversary Prospect 7a , Moscow 117312, Russia
}}\\[.40in]

\end{center}
\begin{abstract}
Distortion of the equilibrium spectra of cosmic neutrinos due to interaction 
with hotter electrons and positrons in the primeval cosmic plasma is considered.
The set of integro-differential kinetic equations for neutrinos is accurately
numerically solved. The relative corrections to neutrino energy densities are
approximately 0.9\% for $\nu_e$ and 0.4\% for $\nu_\mu$ and $\nu_\tau$. This 
effect results in $1.4 \;\cdot 10^{-4}$ increase in the primordial $^4 H \! e$ 
abundance.

\end{abstract}
\newpage
\section{Introduction}

At high
temperatures the primeval cosmic plasma is very close to equilibrium.
This is  
because the reaction rate, $\Gamma = \sigma n$,  typically is much
higher than the expansion rate, $H \sim 1/t$. Here $\sigma$ is the
characteristic cross-section and $n \sim T^3$ is the particle number density.
One can see from the covariant Boltzmann kinetic equation:
\beq{
 (\partial_t - Hp\partial_p) f = I_{coll}
\label{dtf}
}\eeq
that equilibrium for massless particles is not distorted by the expansion of
the universe. Indeed the equilibrium distribution:
\beq{
f_{eq} = [\exp (E -\mu (t) ) / T(t) \pm 1]^{-1}
\label{feq}
}\eeq
which annihilates the
collision integral $I_{coll}$, satisfies this equation if $\dot T /T =-H$
and $\mu (t) \sim T(t)$. However if the particle mass $m$ is nonzero, the
l.h.s. of eq. (\ref{dtf}) cannot vanish for all 
possible values of the momentum 
$p$ with any choice of the functions
$T(t)$ and $\mu (t)$. The deviation from  equilibrium can  roughly be
estimated as $\delta f / f = Hm^2 /(\Gamma TE)$. 

Since photons and electrons are
very strongly coupled, the deviations from  equilibrium in the  
electromagnetic
cosmic plasma are tiny. This is the reason for a perfect Planckian shape
in  the observed spectrum of the cosmic microwave radiation, where the
distortion is known to be smaller than $10^{-4}$.  
One would expect the
same to be true for massless cosmic neutrinos. However this is not so and the
spectral corrections, especially in the high energy tail, 
can be of
several per cent. The physics of the phenomenon is quite simple. At high
temperatures neutrinos and electrons are sufficiently strongly coupled to
each other, they have the equilibrium distributions and their temperatures
are equal. At smaller temperatures, below 2-3 MeV, the neutrinos decouple and
below these temperatures the primeval plasma
consists of two (or better to say four) almost decoupled components, the
electromagnetic and three neutrino ones. In the approximation 
of instantaneous decoupling the
neutrino spectra maintain their  equilibrium shape with the temperature
decreasing as the inverse scale factor, $T_\nu \sim 1/a(t)$.
The temperature of electrons and photons decreases slower because
$e^+ e^-$-annihilation heats up the electromagnetic part of the plasma
when the temperature drops below
$m_e $. Ultimately the ratio of the temperatures reaches the
well known value $T_\gamma /T_\nu = (11/4)^{1/3} = 1.401$, see e.g.
ref. \cite{sw}. This result was obtained under the assumption of entropy
conservation in the electromagnetic sector, which is not necessarily 
true in the non-equilibrium case, where the limiting value of
$T_\gamma /T_\nu$ becomes slightly different. The interactions between the
electromagnetic and neutrino components of the plasma, though weak, are not
completely vanishing and therefore the residual
annihilation $e^+ e^-  \rightarrow \nu \bar \nu$ and, to a lesser extend,
elastic scattering $e\nu \rightarrow e\nu$ slightly heat up the neutrinos and
distort their spectra. The distortions are larger at higher energies because
the weak interactions are getting stronger with rising energy.

The overall
neutrino heating under assumption of equilibrium spectra has been
considered earlier in refs. \cite{dkg,hh,rm}. In this approximation the problem
is very much simplified but the method itself is far from being precise from
the very beginning. A more complete approach with the distortion of the
spectra taken into account has been discussed in several
papers \cite{df,dt,hm} in different approximations and
with somewhat different results. In what follows we present an accurate
numerical treatment of the problem with a better accuracy than that of the
preceding paper \cite{hm} where the numerical solutions of the exact kinetic
equations  were also obtained. Our better accuracy is
achieved by a convenient
reduction of the collision integral to two dimensions and a powerful 
integration method developed in ref. \cite{st}.

\section{Basic Equations}

Instead of time and momenta we choose the following dimensionless variables:
\beq{
 x= m a(t), \,\,  y_j= p_j a(t)
\label{xy}
}\eeq
where $m$ is an arbitrary parameter with dimension of mass,
which we took as $m=1$~MeV and the
scale factor $a(t)$ is normalized so that $a(t) = 1/T_\nu = 1/T_\gamma$ at high
temperatures or at early times. In terms of these variables the kinetic
equation (\ref{dtf}) can be rewritten as:
\beq{
 Hx \partial_x f(x,y_1) = I_{coll}.
\label{hxdxf}
}\eeq
The collision integral $I_{coll}$ is dominated by two-body reactions
$1+2 \rightarrow 3+4$ and is given by the expression:
\begin{eqnarray}
I_{coll} = {1\over 2E_1}\sum \int {d^3 p_2 \over 2E_2 (2\pi)^3}
{d^3 p_3 \over 2E_3 (2\pi)^3}{d^3 p_4 \over 2E_4 (2\pi)^3}
\nonumber \\
(2\pi)^4\delta^{(4)} (p_1+p_2-p_3-p_4) F(f_1,f_2,f_3,f_4)
S\, |A|^2_{12\rightarrow 34}
\label{icoll}
\end{eqnarray}
where $F = f_3 f_4 (1-f_1)(1-f_2)-f_1 f_2 (1-f_3)(1-f_4)$,
$|A|^2$ is the weak interaction amplitude squared summed over spins of
all particles except the first one, and $S$
is the symmetrization factor which includes $1/2!$ for each pair of identical
particles in initial and final states and the factor 2 if there are 2 identical
particles in the initial state; the summation is done over all possible
sets of leptons 2, 3, and 4;
The relevant reactions are presented in Table~1 for the
case when the first particle is
$\nu_e$, and in Table~2 for $\nu_\mu$ or $\nu_\tau$. Our results in these tables
agree with those of ref. \cite{hm} except for the reactions
$\nu_a\nu_a\rightarrow \nu_a \nu_a$ where we took twice larger contribution
because of identical particles in the initial state.

We assume that the
distribution functions for $\nu_\mu$ and $\nu_\tau$ are equal while that for
$\nu_e$ is different because of different strength of charged current
interactions. Similarly we assume that the lepton asymmetry is negligible
so that $f_{\nu} = f_{\bar{\nu}}$. 
Therefore there are two unknown functions of $x$ and $y$:
$f_{\nu_e}$ and $f_{\nu_\mu} = f_{\nu_\tau}$. Each of them satisfies the
kinetic equation (\ref{hxdxf}) with the appropriate collision integral.
We also assume  that the distributions of photons and electrons are the exact
equilibrium ones (\ref{feq}) with the unknown temperature $ T_\gamma (x)$
and zero chemical potential, $\mu(x) = 0$. The
third necessary equation is the covariant energy conservation:
\beq{
x \; {d\rho (x) \over dx } = -3(\rho + P)
\label{drhodx}
}\eeq
where $\rho$ is the total energy density:
\beq{
 \rho = {\pi^2 T^4_\gamma\over 15}  + {2\over \pi^2} \int {dq q^2
\sqrt{q^2 + m^2_e} \over \exp {(E/T_\gamma)} +1 } +
{1\over \pi^2} \int dq q^3 f_{\nu_e} + {2\over \pi^2} \int dq q^3 f_{\nu_\mu}
\label{rho}
}\eeq
and $P$ is the pressure:
\beq{
 P = {\pi^2 T^4_\gamma\over 45}  + {2\over \pi^2} \int {dq q^4 \over
3\sqrt{q^2 + m^2_e}  [\exp (E/T_\gamma) +1 ]} +
{1\over 3\pi^2} \int dq q^3 f_{\nu_e} + {2\over 3\pi^2}
\int dq q^3 f_{\nu_\mu}
\label{p}
}\eeq

The Hubble parameter, $H=\dot a /a$, which enters the kinetic 
equation (\ref{hxdxf}) is expressed through $\rho$ in the usual way,
$3H^2 m^2_{Pl} =8\pi \rho$, ignoring the curvature term and the
cosmological constant.

The collision integral in eq. (\ref{icoll}}) can be reduced from nine to 
two dimensions as
described in Appendix A. For the sake of brevity let us introduce some 
notations: $f_a (p_j) \equiv f_a^{(j)}$,
$d_1 = D_1$, $d_2 (3,4) = D_2(3,4) / E_3 E_4$, and 
$d_3 = D_3/E_1E_2E_3E_4$. The functions $D_j$ are defined in Appendix A.
We do not indicate the arguments for the functions $D_1$ 
and $D_3$ because they are symmetric in all 4 arguments. In terms of these
functions we can write the coupled
kinetic equations for $f_{\nu_e}$ and $f_{\nu_\mu}$ in the following way:
\begin{eqnarray}
 Hx\partial_x f_{\nu_e} ^{(1)} = { G^2_F \over 2\pi^3 p_1} 
\int dp_2 p_2 dp_3 p_3 dp_4 p_4 \delta (E_1 +E_2 -E_3-E_4)
\nonumber \\
\left\{
F[f_{\nu_e}^{(1)},f_{\nu_e}^{(2)},f_{\nu_e}^{(3)},f_{\nu_e}^{(4)}]
[6d_1 - 4d_2(1,4) -4 d_2(2,3) +2d_2(1,2)+2d_2(3,4) +6d_3] \right.
\nonumber \\
+ F[f_{\nu_e} ^{(1)},f_{\nu_\mu} ^{(2)},f_{\nu_e} ^{(3)},f_{\nu_\mu} ^{(4)}]
[4d_1 + 2d_2(1,2) +2d_2(3,4) -2d_2(1,4)-2d_2(2,3) +4d_3]
\nonumber \\
+ F[f_{\nu_e}^{(1)},f_{\nu_e}^{(2)},f_{\nu_\mu}^{(3)},f_{\nu_\mu}^{(4)}]
[2d_1 - 2d_2(2,3) -2d_2(1,4) +2d_3]
\nonumber \\
+ F[f_{\nu_e}^{(1)},f_{e}^{(2)},f_{\nu_e}^{(3)},f_{e}^{(4)}]
[4(g_L^2+g_R^2)(2d_1 - d_2(2,3) -d_2(1,4)+d_2(3,4)
\nonumber \\
+d_2(1,2) +2d_3)
-8g_Lg_R m^2_e (d_1 - d_2(1,3))/E_2E_4]
\nonumber \\   
+ F[f_{\nu_e}^{(1)},f_{\nu_e}^{(2)},f_{ e}^{(3)},f_{e}^{(4)}]
[4g_L^2 (d_1 - d_2(2,3) -d_2(1,4) +d_3)+
\nonumber \\  \left.
4g_R^2 (d_1 - d_2(2,4) -d_2(1,3) +d_3)
+4g_Lg_R m^2_e (d_1 + d_2(1,2))/E_3 E_4]
\right\}\label{fnue}
\end{eqnarray}
and:
\begin{eqnarray}
 Hx\partial_x f_{\nu_\mu} ^{(1)} = { G^2_F \over 2\pi^3 p_1} 
\int dp_2 p_2 dp_3 p_3 dp_4 p_4 \delta (E_1 +E_2 -E_3-E_4)
\nonumber \\
\left\{
F[f_{\nu_\mu}^{(1)},f_{\nu_\mu}^{(2)},f_{\nu_\mu}^{(3)},f_{\nu_\mu}^{(4)}]
[9d_1 - 6d_2(1,4) -6 d_2(2,3) +3d_2(1,2)+3d_2(3,4) +9d_3] \right.
\nonumber \\
+ F[f_{\nu_\mu} ^{(1)},f_{\nu_e} ^{(2)},f_{\nu_\mu} ^{(3)},f_{\nu_e} ^{(4)}]
[2d_1 + d_2(1,2) + d_2(3,4) -d_2(1,4) - d_2(2,3) +2d_3]
\nonumber \\
+ F[f_{\nu_\mu}^{(1)},f_{\nu_\mu}^{(2)},f_{\nu_e}^{(3)},f_{\nu_e}^{(4)}]
[d_1 - d_2(2,3) -d_2(1,4) +d_3]
\nonumber \\
+ F[f_{\nu_\mu}^{(1)},f_{e}^{(2)},f_{\nu_\mu}^{(3)},f_{e}^{(4)}]
[4(\tilde{g}_L^2+g_R^2)(2d_1 - d_2(2,3) -d_2(1,4)+d_2(3,4)
\nonumber \\
+d_2(1,2) +2d_3)
-8\tilde{g}_Lg_R m^2_e (d_1 - d_2(1,3))/E_3 E_4]
\nonumber \\   
+ F[f_{\nu_\mu}^{(1)},f_{\nu_\mu}^{(2)},f_{ e}^{(3)},f_{e}^{(4)}]
[4\tilde{g}_L^2 (d_1 - d_2(2,3) -d_2(1,4) +d_3)+
\nonumber \\  \left.
4g_R^2 (d_1 - d_2(2,4) -d_2(1,3) +d_3)
+4\tilde{g}_Lg_R m^2_e (d_1 + d_2(1,2))/E_3 E_4]
\right\},\label{fnumu}
\end{eqnarray} 
where $g_{L} =  1/2 + \sin^{2} \theta _{W}$, 
$g_{R} = \sin^{2} \theta _{W}$ and $\tilde{g}_{L} = - 1/2  +
\sin^{2} \theta _{W}$,
 $f_e = [\exp( E/T_\gamma ) +1 ]^{-1} $, and the temperature $T_\gamma$ is
determined from eq. (\ref{drhodx}). Thus we have the complete 
system of three equations for three unknown functions $T_\gamma (x)$,
$f_{\nu_e} (x,y)$, and  $f_{\nu_\mu} (x,y)$ which we solve numerically.
The solution has been found in two different but equivalent ways. First, the 
system has been solved directly, as it is, for the full distribution 
functions $f_{\nu_j}(x,y)$
and, second, for the small deviations $\delta _{j}$ from equilibrium 
$f_{\nu_j} (x,y) = f_{\nu_j}^{(eq)} (y) (1 +\delta_j (x,y))$, where 
$f_{\nu_j}^{(eq)} = [\exp( E/T_\nu  ) +1 ]^{-1}$ with $T_\nu = 1/a$.
In both cases the 
numerical solution has been done exactly, not perturbatively.  
So with infinitely good numerical precision the results must be the same.
However since the precision is finite, different ways may give 
different results
and their consistency is a good check of the accuracy of the calculations.
It is convenient to introduce $\delta (x,y)$ because the dominant terms
in the collision integrals, containing neutrinos only, cancel and 
subdominant terms are proportional to $\delta$. In the parts of the
collision integrals 
which contain electrons, there are also contributions proportional to the 
difference in temperatures $(T_\gamma - T_\nu)$. We have found that the 
results of these two methods, i.e. solving either for $f_\nu$ or $\delta$,
are very close (see the next Section and Table 3).

To check possible algebraic errors we have made the following procedure. 
First, we have solved numerically the three relevant equations in the Boltzmann 
approximation, simply changing the function $F$ which is defined after 
eq. (\ref{icoll}) to $F\rightarrow f_3 f_4 - f_1 f_2$ with the same 
coefficients and d-functions in the collision integral
as presented above. Second, we repeated the
numerical calculation but this time we changed the terms describing the direct 
reactions (the ones that enter $I_{coll}$ with negative signs) with 
(presumably) identical terms where the integration over the phase space of the
final particles had been made explicitly. That is, the positive 
terms were  
described through d-functions as above, while negative ones had been calculated in a
different and independent way. The expressions for the negative terms  
are presented in 
Appendix~B. The results of these two numerical calculations are in perfect 
agreement. This excludes possible accidental errors in calculations of
D-functions (see Appendix A) as well as in the typing of 
the program for the numerical integration.

\section{Description of Calculations and Results}

\subsection{ Initial conditions}

We have numerically calculated the evolution of the distribution
functions satisfying the system of eqs. (\ref{drhodx},\ref{fnue},\ref{fnumu})
in terms of the dimensionless time $x = m_0 a$ in the interval 
$x_{in} \le x \le x_f$, where $a$ is the scale factor.
With our normalization the "time" $x= 1$ corresponds to the  
neutrino temperature $T_\nu = m_0=1 MeV$.\footnote{Strictly speaking  
for non-equilibrium neutrinos we 
cannot introduce temperature $T_{\nu}$. But since 
non-equilibrium corrections to the neutrino spectrum are small (see below)
we  sometimes use the notation $T_{\nu}$ instead of $1/a$, because 
in equilibrium $T_\nu a =1$.}

For large values of $x$ the collision integrals in the 
r.h.s. of eqs. (\ref{fnue},\ref{fnumu}) 
are suppressed by the factor $1/x^2$.
We have found that at $x \approx 50$  all the functions
$f_{\nu_e}$, $f_{\nu_{\mu(\tau)}}$, and $T_\gamma$ reach
their asymptotic values, so to be certain that we are in the asymptotic limit 
we have chosen the final time $x_f = 60$. 
For the initial time $x_{in}$ we have chosen three different values 
$x_{in}=0.1,\; 0.2, \;{\rm and} \; 0.5$ which  correspond
to $T_{\nu}=10 MeV,\; 5 MeV,\; {\rm and} \; 2 MeV$ respectively.    
In all the cases we have assumed that  the neutrinos are in thermal
equilibrium for  $x \leq x_{in}$  with the distribution functions
$f_{\nu_j}^{(eq)} = 1/[\exp(y) +1 ]$.

For the dimensionless momentum $y = pa$ we took a $100$ and a $200$ point 
grid, equally spaced in the
region $0 \le y \le 20$. In contrast to  ref. \cite{hm} 
we have included the point $y=0$ in which the collision integrals
in eqs. (\ref{fnue},\ref{fnumu}) have different analytical expressions due to 
the cancellation of the factor $1/p_1^2$ in front of the collision integrals by 
the corresponding factors in the functions $D_i$. 
This allows us to compare the 
collision integral at $y=0$ with the integral in a nearby point $0 < y \ll 1$ 
in order to check that the    numerical errors  are small 
in the region of small momenta $y<1$.

\subsection{Numerical results}

There are three phenomena which play essential roles in the
evolution of the neutrino distribution functions. The first of 
them is the temperature difference between photons and $e^\pm$ on one side
and neutrinos on the other, which arises due to the 
heating of the electromagnetic 
plasma by $e^+ e^- $-annihilation. 
Through the interactions of neutrinos 
with electrons this temperature difference
leads to non-equilibrium distortions of the neutrino spectra.
The temperature difference is essential in the interval 
$1<x<30$. The second effect is the freezing of the neutrino interactions
because the collision 
integrals in the r.h.s. of eqs. (\ref{fnue},\ref{fnumu}) drop as 
$1/x^2$. At small $x \ll 1$ the collisions are very efficient but at 
$x>1$ they are strongly suppressed.
The third important phenomenon is the elastic $\nu \nu$-scattering 
which smoothes  down the non-equilibrium corrections to the neutrino 
spectrum. It is especially important at small $x<1$.

Accordingly, the evolution of the distribution functions proceeds as follows.
At $x<0.2$ the temperature difference is negligibly small while
$\nu \nu$-scattering is strong. Therefore the neutrino distribution 
functions are practically in equilibrium. 
In the interval $0.2 \le x \le 4 $ all three effects are essential and
during this period the neutrino spectra are  distorted and simultaneously
the photon temperature slightly drops down because of the energy transfer 
from $e^\pm$ to neutrinos, as compared to the photon temperature in the
approximation where the neutrino decoupling is instantaneous.
When $x>4$ the  collision integrals in the r.h.s. of 
eqs. (\ref{fnue},\ref{fnumu}) are small and the temperature difference 
practically does not affect the neutrino distribution functions, even though
the temperature difference is the largest in this time interval.

In Figs. 1-5 we present the results of the numerical solution  
of the system of eqs. (\ref{drhodx},\ref{fnue},\ref{fnumu})  
in the interval $0.1 \le x \le 60$  (approximately 4000 points in  $x$) 
both for   Fermi-Dirac (FD) and  Maxwell-Boltzmann (MB) statistics. 
In these runs we used a 100 point grid in the momentum
interval $0 \le y \le 20$.

The ratio $T_\gamma/T_\nu$  as a function of $x$ is presented 
in Fig. 1 for FD and MB statistics. If we neglect the exchange of energy 
between $e^\pm$ and 
neutrinos in eq. (\ref{drhodx}) (which is equivalent to 
entropy conservation) the asymptotic values are  
$T_\gamma/T_\nu=1.4010$ for FD statistics and $T_\gamma/T_\nu=1.4423$ for
MB statistics. When the energy exchange is taken into account
(energy conservation) these ratios have
slightly different values  $T_\gamma/T_\nu=1.3991$ (or 1.3994, using the full
distribution function) for FD statistics 
and $T_\gamma/T_\nu=1.4404$ for MB statistics. 

In Fig. 2 the correction to the total neutrino energy density
 $\delta\rho_\nu /\rho_{ eq}$  for the case of FD statistics is plotted.
 Here $\rho_{eq} = (7\pi^2 / 240) (m/x)^4$ is the unperturbed
energy density of neutrinos and:
\begin{equation}
\delta\rho_\nu=\frac{ \rho_{\nu_e} + 2  \rho_{\nu_\mu}}{3} - \rho_{ eq} ~, 
\label{drhonu}
\end{equation}   
with  $\rho_{\nu_e} $ and $\rho_{\nu_\mu}$ found from our numerical solution.
The upper curve in Fig. 2 corresponds to the entropy conservation case, 
while the lower one corresponds to the energy conservation case.
The effect is smaller in the second case because the "driving force", 
which is proportional to the temperature difference, is lower due to 
the excessive photon cooling. We can see that while 
the difference in the asymptotic values of 
$T_\gamma/T_\nu$ for these two cases
is as small as $0.14\%$  the difference in $\delta\rho_\nu / \rho _{eq}$ is
as large as $20\%$. 
This can be explained in the following way.
Compare Fig. 1 and Fig. 2 for the FD case. We see that already at 
$x=4$ the value of the function $\delta\rho_\nu / \rho _{eq}$ is  close to its 
asymptotic value. At the same
moment, $x=4$, the ratio $T_\gamma/T_\nu$ has a value of $1.08$ which is much
smaller than the asymptotic one, 1.401. 
This means that the dominant contribution
to the distortion of the neutrino spectra comes from the period when the
temperature difference $\Delta T = T_\gamma - T_\nu$ was rather small. The
smallness of $\Delta T$ was compensated by a more efficient energy exchange
between $\nu$ and $e^{\pm}$ at smaller $x$. In this range of $x$, a relatively
small variations in $T_\gamma /T_\nu$ connected to different approximations
made in the calculations, should be compared with the small difference 
$\Delta T$. Thus a small variation in  $T_\gamma /T_\nu$ is
relatively enhanced, and we see how these small 
variations can produce the strong 20\% effect.
 
In Fig. 3 the relative deviations of the neutrino spectra from
the equilibrium one, $\delta_{j}(x,y)=(f_{\nu_j}-f_{eq})/f_{eq}$, are 
presented as
functions of  $x$ for several  
values of the momentum $y=3,5,7$. In Fig. 3a we present the results for 
electronic neutrinos and in Fig. 3b for muonic (tau) ones. 
For large momenta $y$ the deviation from the equilibrium is larger and
the maximum asymptotic value is reached later (at larger $x$). 
This is a result of
stronger interactions of more energetic neutrinos.
 
In Fig. 4 we compare the deviations  
from the equilibrium distributions, $\delta_{\nu_e}$ and
$\delta_{\nu_{\mu(\tau)}}$ for FD and MB statistics.  
We plot $\delta_i$ for the same  value 
of the  momentum $y=5$ as  functions of $x$. We see that the results 
for the case of Boltzmann statistics are larger than those for the 
Fermi one by approximately $25\%$.  For both FD and
MB statistics the spectral distortion for $\nu_e$ is more than twice larger
than that for $\nu_\mu$ or $\nu_\tau$. This is due to a stronger coupling of
$\nu_e$ to $e^\pm$.
 
In Fig. 5 we plot the asymptotic values of the  
corrections to the neutrino distributions 
$\delta_j = (f_{\nu_j} - f_\nu^{eq})/f_\nu^{eq} $
as functions of the dimensionless momentum $y$.
The dashed lines $a$ and $c$ correspond to Maxwell-Boltzmann statistics
and the solid lines $b$ and $d$ correspond to Fermi-Dirac statistics.
The upper curves $a$ and $b$ are for electronic neutrinos and the lower 
curves $c$ and $d$ are for muonic (tau) neutrinos. All the curves can be well 
approximated by a second order polynomial 
in $y$ like $\delta = A y (y -B)$ in agreement with ref. \cite{df}.

\subsection{ Checking the results}

First, let us discuss the choice of the initial time $x_{in}$. We made 
the runs for the system of eqs. (\ref{drhodx},\ref{fnue},\ref{fnumu}) with
three different values $x_{in}=0.1, 0.2$ and $0.5$. We found that the
results of the runs with $x_{in}=0.1$ and $x_{in}=0.2$ are the same with the
accuracy of $10^{-5}$ in the distribution function for all 
values of $x$ (see Table 3). This 
means that for $x \le 0.2$ we can neglect the non-equilibrium corrections 
to the neutrino distribution functions. For larger $x$ this is not so and e.g.
the spectral distortion calculated with
$x_{in}=0.5$ is smaller by approximately $15\%$ than the one calculated with
$x_{in}=0.2$.

In order to check the errors connected with a finite number of points in 
the momentum interval $0 \le y \le 20$ we took a 100 and a 200 point grids. 
We found that the difference in the results is small, which means that 
a 100 points grid is a good 
approximation for the numerical solution of our system of  
kinetic equations.

The errors connected with a finite number of points
in time $x$ are much more important. 
We control these errors in the following way. First, we run the program
with some fixed number of points in time $x$, distributed in the time interval
$x_{in}<x<x_f$ in such a way that the distribution functions do not change
significantly at any 
momentum point $y$ during one time iteration $dx$. Then we
run the 
program for the entropy conservation law (i.e. with equilibrium neutrinos)
with the same values of time $x_i$ as in the first run. Then we compare
the asymptotic values of the temperature ratios with the 
sample values which are   
$T_\gamma/T_\nu=(11/4)^{1/3} = 1.4010$ for FD statistics and 
$T_\gamma/T_\nu={3}^{1/3} = 1.44225$ for
MB statistics. We found that when we  run with  400 time points, the
numerical error in these temperature ratios are of the order $ \sim 0.001$, 
and this  value is only one order of magnitude smaller than 
the effect itself, which is not good enough. 
However,  in the case of 4000
time points this error is as small as  $\sim 10^{-4}$. In our calculations
we used a 4000 point grid in time.  

We also checked the errors related to the finite number of points in time 
in a different way.
Instead of the simple time evolution we used the Bulirsch-Stoer method, 
described in the book \cite{numrec}. We found, that this method
gives the same results as the simple time evolution 
with 4000 time points (see the first two rows in Table 3).

Besides reducing the numerical errors we have used two  different analytical 
approaches in order to check  the results. 
In the case of FD statistics  we solve the system of 
eqs. (\ref{drhodx},\ref{fnue},\ref{fnumu}) both for the full neutrino
distribution functions  $f_{\nu _j}$ and for the deviations from equilibrium 
$\delta_j = (f_{\nu_j}-f_\nu^{eq})/f_\nu^{eq} $. In the last case 
the contributions to the collision integrals for
all the processes vanish for vanishing $\delta$, except for the
interactions of neutrinos with electrons, where the "driving force" term,
proportional to the temperature difference between $\nu$ and $e^\pm$, gives
a nonzero contribution. 
Comparing the results in these two 
cases (with the same
values of all parameters) we have found that the non-equilibrium corrections 
to the neutrino spectra are systematically  smaller by about 10\%  when the
equations are solved for the full distribution functions $f_{\nu_j}$. 
The comparison of the results of the different ways of calculation are 
summarized in Table 3.

\subsection{Helium abundance}

We have modified the standard nucleosynthesis code 
(ref. \cite{kawano}) in the following way.
The neutrino distribution functions have been parametrized as:
\begin{equation}
f _{\nu_{i}} \rightarrow f_{\nu_{i}} \cdot (1 + \delta_i (p,T_{\nu})),
\label{corrdist}
\end{equation}
where $\delta_i$ were fitted as functions of momentum and temperature in
accordance with our numerical solutions.

First, the neutrino energy densities $\rho_{\nu_{i}}$ 
have been calculated with the 
corrected spectra. This changes the universe cooling rate and has 
the major impact on the helium abundance. 

Second, the 6 weak interaction rates for $(n \leftrightarrow p)$-reactions
have been modified with the account for the spectral distortion of the
electronic neutrinos.
The 3 reactions with $\nu_e$ or $\bar{\nu}_e$ in the initial state, 
$n \; \nu_e \rightarrow p \; e^{-}$, $p \; \bar{\nu}_e  \rightarrow 
n \; e^{+}$ and  $p \; \bar{\nu}_e \; e^{-} \rightarrow  n $, are more 
important
and the neutrino distribution functions have been corrected 
according to eq.~(\ref{corrdist}).
However, for the 3 inverse reactions with neutrinos in the final state,
$n \; e^+ \rightarrow p \; \bar{\nu}$, 
$n \rightarrow p \; e^- \; \bar{\nu}$ and 
$p \; e^- \rightarrow n \; \nu$, the spectral modification results 
only  in a small correction to the Pauli
suppression factor,  $(1 - f_{\nu}) \rightarrow
(1 - f_{\nu} \cdot (1 + \delta))$, and the effect of the corrections to
these last 3 reactions is negligible.

Finally the law of the evolution of the photon temperature is slightly changed,
which in turn also affects the reaction rates.
The correct photon temperature is calculated by 
including the neutrino  energy densities into the equation governing the
variation of the photon temperature in the course of the universe expansion:
\begin{equation}
\frac{dT_{\gamma}}{d \; lnV} = - \frac{\rho _{EM} + p _{EM} 
+ 4/3 \; \rho _{\nu}}{d \rho_{EM} / dT_{\gamma} +d \rho_{\nu} / dT_{\gamma}},
\label{phottempofexpansion}
\end{equation}
where $V = a^3$. If neutrinos had  the equilibrium distribution
their energy density would satisfy the covariant conservation law 
(\ref{drhodx})
separately so that their contribution into 
eq. (\ref{phottempofexpansion}) would cancel out. However this is not true if
the energy exchange between $\nu$'s and $e^\pm$ is taken into account.

The final change in helium-4 abundance has been found to be
$\Delta Y = 1.4 \cdot 10^{-4}$. Similarly we find relative changes in
the $^3 H \! e$ and $^7 Li$ abundances of the order 
$5  \cdot 10^{-4}$ and $-6  \cdot 10^{-4}$  respectively.

\section{Discussion}

Our results agree reasonably well with the previous numerical calculations of
ref. \cite{hm}. 
For example we obtain $\delta \rho_{\nu_e} /\rho_{\nu_e} = 0.94\%$ (0.83\%)
and $\delta \rho_{\nu_\mu} /\rho_{\nu_{\mu}} = 0.40\%$ (0.33\%) (see Table 3),
to be compared with 0.83\% and
0.41\% respectively obtained in ref. \cite{hm}. 
For the shape of the spectral distortion the difference is larger, 
especially for low $y$. To characterize the
difference let us introduce the effective
neutrino temperature as $T_{eff} = E /\ln [(1-f)/f]$, which we 
compare to the unperturbed  temperature $ T_{0}$ 
calculated in the limit of entropy 
conservation i.e.  in neglect of the neutrino heating. The difference:
\beq{
{T_{eff} - T_0  \over T_0 } = {1  - e^{-y} \over y}\, \delta (x,y)
\label{teff}
}\eeq
goes to a small negative value for small $y$ in 
our case and to a considerably larger
positive one according
to ref. \cite{hm}. We ascribe this difference to a better accuracy of our 
calculations which is most profound at low $y$
(in particular we took a 100-200 point grid in comparison to 35 in 
ref. \cite{hm}). The physical reason for this negative result is the elastic
$\nu e^{\pm}$-scattering. 
Since electrons and positrons have  a slightly higher temperature 
than the neutrinos, the scattering will depopulate the low momentum region 
pushing neutrinos to higher momenta. 
Our agreement with ref. \cite{hm} is much better for the physically 
more important
region of high $y$. The distribution function decreases exponentially,
whereas the correction $\delta$ increases quadratically. Moreover the 
weak interaction rates are typically proportional to the energy squared,
so the dominant
contribution of $\delta$ to neutron-proton reactions lays near $y = 5$. 
For large $y$
our results are systematically larger than that of ref. \cite{hm}
by approximately 15\%. This difference cannot be ascribed to  
the difference by a factor two in the collision integrals for the
scattering of identical neutrinos because that 
leads to the opposite sign of the
effect and is rather small, giving corrections of a few per cent. 
We believe that the difference between us and ref. \cite{hm} arises 
because of different numerical accuracies. 

Another source of the difference in the results is the different treatment of
the cooling of the electromagnetic plasma. The simplest approach is to 
neglect the loss of photon energy which goes into neutrinos. 
In this case the entropy of photons 
and $e^+e^-$-pairs is conserved and the function $T_{\gamma}(x)$ can 
be determined
from the relation $x^3(\rho_{em} + P_{em} )/T_\gamma = const$. In this
approximation the cooling of the electromagnetic plasma is induced only by
the expansion of the universe. The energy transmitted to neutrinos results in
an excessive cooling which  we have calculated exactly from the 
more complicated 
equation (\ref{drhodx}), which does not necessarily imply conservation of 
entropy in the non-equilibrium case. 
The effect is found to be significant. 
Inclusion of the additional cooling of photons  amplifies 
the neutrino spectral distortion by approximately 20\% (see fig. 2).
This can be explained
as follows. The distortion of the spectrum is determined by the temperature
difference $\Delta T / T_{\nu} = (T_\gamma - T_\nu) / T_\nu$ and 
by the strength 
of neutrino-electron interactions. The former rises with the expansion 
while the latter dies down. At large $x$ the temperature difference $\Delta T$ 
is much larger than the correction to the photon temperature but at smaller
$x = O(1)$ they do not differ so much.
The dominant contribution to the distortion of the neutrino spectrum
comes just from the period when $x$ is near unity and $\Delta T$ is relatively 
small, so that the small correction to $T_\gamma$ is relatively enhanced.  
This effect was neglected in the previous papers \cite{df,hm} so their 
results are somewhat underestimated; it was approximately  calculated in
ref. \cite{dt} under the assumption that the 
neutrinos have equilibrium spectra.
However the spectral modification gives a contribution into additional photon
cooling of the same order of magnitude and should be taken into account.
 
In the earlier papers \cite{df,dt} the effect was considered in the Boltzmann
approximation which simplifies the calculations very much. Another simplifying 
assumption, previously used, is the neglect of the electron mass in the
collision integrals for $\nu e $-scattering and for annihilation
$\bar \nu \nu  \rightarrow e^+e^-$. We have also made numerical calculations in
these approximations and found that both of these give rise to a larger 
spectral distortion in agreement with ref. \cite{hm}. 
In ref. \cite{dt} the effect 
was calculated numerically while in ref. \cite{df} an approximate analytical
expression was derived. However in ref. \cite{df} the influence of 
the back-reactions which  
smoothes down the spectral distortion was underestimated due to a numerical
error in the integral. With the correction of this error the effect
should be twice smaller (in the approximations of that paper). Our numerical
results in the limit of Boltzmann statistics and for $m_e = 0$ are in a 
reasonable agreement with the corrected estimate of that paper: 
\beq{
 \delta_{\nu_e} \approx 3\times 10^{-4} y (11y/4 -3).
\label{deltanue}
}\eeq


{\bf Acknowledgment.}
The work of AD and SH was supported in part by the Danish National Science 
Research Council
through grant 11-9640-1 and in part by Danmarks Grundforskningsfond through its
support of the Theoretical Astrophysical Center. 
The work of DS was supported in part 
by the Russian Foundation for Fundamental Research through grant 95-02-04911A
and by the INTAS program of EC through grant INTAS-93-1630 (extension).
DS thanks the Theoretical Astrophysical Center for hospitality during the last 
stages of this work.

\bigskip
\appendix

\section{Reduction of the collision integral}
\label{sec:Simp}
In this Appendix we will make analytically as many integrations 
in the collision integral as  possible. 
In the case we are interested in, {\it i.e.}
four-particle interaction vertex and the isotropic one-particle distribution
functions, seven out of nine integrations can be done and only two integrals
upon the momentum remain for numerical treatment. We
start from the general form of the collision integral in the 
space-homogeneous case:
\begin{equation}
I_{\mbox{coll}} = \frac{1}{2E_1} \int
(2\pi)^4\delta^4(\sum_i p_{\mu i}) |M_{fi}(|{\bf p} |)|^2  F(f) \prod_{i=2}^{i=4}
\frac{d^3{\bf p}_i}{(2\pi)^3 2E_i}~,
\label{Icoll}
\end{equation}
where the particle energy is $E_i=\sqrt{m_i^2+{\bf p}_i^2}$ 
and $F(f)$  is defined as:
\begin{equation}
F(f)=[1-f_1][1-f_2]f'_1f'_2-[1-f'_1][1-f'_2]f_1f_2,
\label{Fun1}
\end{equation}
where the distribution functions $f_j$ only depend upon the moduli 
of the particle momentum  and time $f_j (|{\bf p} |,t)$. 
We make use of the identity:
\begin{equation}
\delta^3(\sum {\bf p}_i)=
\int  e^{i({\bf \lambda},{\bf p}_1+{\bf p}_2-
{\bf p}'_1-{\bf p}'_2)}\, \frac{d^3 {\bf \lambda}}{(2\pi)^3 }
\label{delta}
\end{equation}
and explicitly separate out the angle integrations:
\begin{equation}
\nonumber
d^3p_i = d\phi_i d\cos\theta_i p_i^2dp_i \equiv  p_i^2 dp_i d\Omega_i
~.
\end{equation}
The collision integral in eq. (\ref{Icoll}) takes the form:
\begin{equation}
I_{\mbox{coll}} = \frac{1}{64\pi^3E_1 p_1} \int
\delta(E_1+E_2-E'_1-E'_2)  F(f)
D(p_1,p_2,p'_1,p'_2) \frac{p'_1dp'_1}{E'_1} 
\frac{p'_2dp'_2}{E'_2}
\frac{p_2dp_2}{E_2} ~.
\label{kin1}
\end{equation}
Here we have defined:
\begin{eqnarray}
D(p_1,p_2,p'_1,p'_2) &\equiv& \frac{p_1 p_2 p'_1 p'_2}{64 \pi^5}
\int_0^{\infty} \lambda^2 d\lambda
\int e^{i({\bf p}_1,{\bf \lambda})}d\Omega_\lambda 
\int e^{i({\bf p}_2,{\bf \lambda})}d\Omega_{p_2} 
\nonumber\\ 
&&
\int e^{-i({\bf p}'_1,{\bf \lambda})}d\Omega_{p'_1} 
\int e^{-i({\bf p}'_2,{\bf \lambda})}d\Omega_{p'_2} |M_{fi}(|{\bf p} |)|^2~.
\label{V}
\end{eqnarray}

In the 
four-fermion approximation all the possible squared matrix elements consist
of two kinds of terms:
\begin{equation}
K_1( {q_1}_\mu q_2^\mu)({q_3}_\nu q_4^\nu) = K_1(E_1E_2 - {\vec q_1}{\vec q_2})
(E_3 E_4 - {\vec q_3}{\vec q_4})
\label{m1}
\end{equation}
and: 

\begin{equation}
K_2m^2( {q_3}_\mu q_4^\mu) = K_2m^2(E_3E_4 - {\vec q_3}{\vec q_4})~,
\label{m2}
\end{equation}
where every $q_i$ corresponds to some of $p_i$.
In order to perform the angle integrals we define the 
angles between ${\vec q_i}$
through angles from eq. (\ref{V}) as: 
\begin{equation}
\cos\phi_{12} = \sin\theta_1 \sin\theta_2 \cos(\varphi_1-\varphi_2)
+\cos\theta_1 \cos\theta_2~.
\label{teta}
\end{equation}
The integration in the angles $\varphi_1$  or $\varphi_2$ 
over the total period  and the structure of the matrix 
elements in eqs. (\ref{m1},\ref{m2}) result in the vanishing of
the first term in eq. (\ref{teta}).  Then all 
angle integrals in eq. (\ref{V}) can be trivially taken
 and we come to three kinds of 
integrals\footnote{In the case that one of $q_3$ or $q_4$
in the second integral $D_2$ corresponds to the incoming particle and 
the other to the outgoing one, $D_2$
changes sign.}:
\begin{equation}
D_1 = \frac{4}{\pi}\int_0^\infty \frac{d\lambda}{\lambda^2}
\sin(\lambda q_1) \sin(\lambda q_2) \sin(\lambda q_3)  \sin(\lambda q_4)~,
\label{D1}
\end{equation}
\begin{equation}
D_2(3,4) = \frac{4 q_3 q_4}{\pi}\int_0^\infty \frac{d\lambda}{\lambda^2}
\sin(\lambda q_1) \sin(\lambda q_2) 
\left[\cos(\lambda q_3)-\frac{\sin(\lambda q_3)}{\lambda q_3}\right]  
\left[\cos(\lambda q_4)-\frac{\sin(\lambda q_4)}{\lambda q_4}\right]  
~,
\label{D2}
\end{equation}
and:
\begin{eqnarray}
D_3 &=& \frac{4 q_1q_2q_3q_4}{\pi}\int_0^\infty \frac{d\lambda}{\lambda^2}
\left[\cos(\lambda q_1)-\frac{\sin(\lambda q_1)}{\lambda q_1}\right]  
\left[\cos(\lambda q_2)-\frac{\sin(\lambda q_2)}{\lambda q_2}\right]  
\nonumber \\ &&
\left[\cos(\lambda q_3)-\frac{\sin(\lambda q_3)}{\lambda q_3}\right]  
\left[\cos(\lambda q_4)-\frac{\sin(\lambda q_4)}{\lambda q_4}\right]  
\label{D3}
\end{eqnarray}
For the matrix element squared given by eq. (\ref{m1}) the function $D$ 
in eq. (\ref{V}) can be written in the form:
\begin{equation}
D = K_1 (E_1 E_2 E_3 E_4 D_1 +D_3) + 
K_1 \left[E_1 E_2 D_2(3,4)+E_3 E_4 D_2(1,2)\right]
~, 
\label{con1}
\end{equation}
while for the matrix element squared given by eq. (\ref{m2}) we  
get:
\begin{equation}
D = K_2 E_1 E_2\left[ E_3 E_4 D_1  + D_2(3,4)\right]
  ~.   
\label{con2}
\end{equation}

The calculation of the integrals (\ref{D1}-\ref{D3}) is straightforward.
The functions $D_1$ and $D_3$ are symmetric with respect to permutations of
any variables and $D_2$ is symmetric under permutations $1\leftrightarrow 2$
and $3\leftrightarrow 4$. In what follows we present 
the expression for $D_2(3,4)$.
Without loss of generality we can assume that $q_1>q_2$ and $q_3>q_4$.
Then we will have four different cases, which depend upon relations between
the momenta.\footnote{In the general case there are  two other
possibilities  
$q_1>q_2+q_3+q_4$ and $q_3>q_1+q_2+q_4$. However they are unphysical and
give zero answer
for all integrals, $D_1=D_2=D_3=0$. }
\vskip0.5cm
a) $q_1 + q_2 > q_3 + q_4$ and $q_1+q_4 > q_2+q_3$,

\begin{equation}
D_1 = \frac{1}{2} (q_2+q_3+q_4-q_1)
\label{D1a}
\end{equation}
\begin{equation}
D_2 = \frac{ 1}{12}((q_1-q_2)^3 + 2(q_3^3+q_4^3) -3(q_1-q_2)
(q_3^2+q_4^2))
~,
\label{D2a}
\end{equation}
\begin{eqnarray}
D_3 &=& \frac{1}{60}
  \left(q_1^5 - 5q_1^3q_2^2 + 5q_1^2q_2^3 - q_2^5 
\right.
\nonumber \\ &-&
5q_1^3q_3^2 + 
5q_2^3q_3^2 + 5q_1^2q_3^3 + 5q_2^2q_3^3 - q_3^5 - 5q_1^3q_4^2
\nonumber \\ &+&
\left. 
  5q_2^3q_4^2 + 5q_3^3q_4^2 + 
5q_1^2q_4^3 + 5q_2^2q_4^3 + 5q_3^2q_4^3 - q_4^5\right)
\label{D3a}
\end{eqnarray}

\vskip0.5cm

b) $q_1 + q_2 > q_3 + q_4$ and $q_1+q_4 < q_2+q_3$,

\begin{equation}
D_1 =  q_4
\label{D1b}
\end{equation}
\begin{equation}
D_2 = \frac{ 1}{3}q_4^3
~,
\label{D2b}
\end{equation}
 
\begin{equation}
D_3 = \frac{1}{30}q_4^3
  \left(5q_1^2+5q_2^2 + 5q_3^2 - q_4^2 \right) 
\label{D3b}
\end{equation}

\vskip0.5cm
c) $q_1 + q_2 < q_3 + q_4$ and $q_1+q_4 < q_2+q_3$,

\begin{equation}
D_1 = \frac{1}{2} (q_1+q_2+q_4-q_3)
\label{D1c}
\end{equation}
\begin{equation}
D_2 = \frac{ 1}{12}(-(q_1+q_2)^3 - 2q_3^3+2q_4^3 +3(q_1+q_2)
(q_3^2+q_4^2))
~,
\label{D2c}
\end{equation}
 
$D_3$ is the same as eq. (\ref{D3a}) 
but with the change
of  variables 
$1 \leftrightarrow 3$ and $2 \leftrightarrow 4$.
\vskip0.5cm

d) $q_1 + q_2 < q_3 + q_4$ and $q_1+q_4 > q_2+q_3$,

\begin{equation}
D_1 =  q_2
\label{D1d}
\end{equation}
\begin{equation}
D_2 = \frac{1}{6}q_2\left( 3q_3^2+3q_4^2-3q_1^2-q_2^2\right)
~,
\label{D2d}
\end{equation}
 
\begin{equation}
D_3 = \frac{1}{30}q_2^3
  \left(5q_1^2+5q_3^2 + 5q_4^2 - q_2^2 \right) 
\label{D3d}
\end{equation}

Now, the energy  $\delta$-function can be trivially integrated away and we are
left with the following 
two-dimensional integral upon energies of incoming particles:

\begin{equation}
\frac{df (E_1,t)}{dt} = \frac{1}{64\pi^3E_1p_1} \int \int
  F(f) (\sum_kD_k) \frac{p'_1dp'_1}{E'_1} 
\frac{p'_2dp'_2}{E'_2}~,
\label{kin2a}
\end{equation}
where $E_2 = E'_1+E'_2-E_1 $,
$p_2=\sqrt{E_2^2-m_2^2}$
and the summation in eq. (\ref{kin2a}) is made
over all $D_k$ contributing to the
squared matrix elements in eqs. (\ref{m1},\ref{m2}). 
In the integration one should
take into account the 
 $\theta$-functions  corresponding to the cases a)-d) as well 
as the $\theta$-function related to the energy conservation, 
$\theta (E_3+E_4-E_1-m_2)$.

We can further check our calculations of the D functions, 
using the following trick. In the
case $m_1^2+m_2^2=m_3^2+m_4^2$ the
l.h.s. of the eq. (\ref{m1}) can be rewritten as:

\begin{equation}
K_1( {q_1}_\mu q_2^\mu)({q_3}_\nu q_4^\nu) = K_1( {q_3}_\mu q_4^\mu)^2~.
\label{m1a}
\end{equation}
We can now integrate the r.h.s. of eq. (\ref{m1a})  over angles.  
Consequently we 
obtain  integrals in the form of eqs. (\ref{D1},\ref{D2}) 
and one more integral:

\begin{eqnarray}
D_4(3,4) &=& \frac{4 q_3^2q_4^2}{\pi}\int_0^\infty \frac{d\lambda}{\lambda^2}
\sin(\lambda q_1) \sin(\lambda q_2)  
\left[\sin(\lambda q_3) -\frac{2}{\lambda q_3}\left(
\cos(\lambda q_3)-\frac{\sin(\lambda q_3)}{\lambda q_3}\right)\right]
\nonumber \\ &&  
\left[\sin(\lambda q_4) -\frac{2}{\lambda q_4}\left(
\cos(\lambda q_4)-\frac{\sin(\lambda q_4)}{\lambda q_4}\right)\right]    
\label{D4}
\end{eqnarray} 

The  equivalence of the two approaches requires that the following
two conditions should be fulfilled:
\begin{equation}
D_2(1,2) \equiv \frac{p_1^2+p_2^2-p_3^2-p_4^4}{2} D_1 + D_2(3,4)  ~ ,
\label{rel1}
\end{equation}
 and 
\begin{equation}
D_3 \equiv \frac{p_1^2+p_2^2-p_3^2-p_4^4}{2} D_2(3,4) + D_4(3,4) ~.
\label{rel2}
\end{equation}
We have checked that this is indeed true, and 
thus assured that the D functions have been calculated correctly.
\bigskip
 
\section{The collision integrals in the Boltzmann approximation}

We have also checked the correctness of our calculations in the
following way. In the Boltzmann approximation we have calculated
everything twice, first letting the computer
solve  the equations directly, and second, by using different exact expressions
in a half of the collision integral. In the Boltzmann approximation
we reduced the first half of the collision integrals to the following
one dimensional integrals over momentum.

For the reactions $\nu _{e} \nu _{e} \rightarrow \nu _{e} \nu _{e} $ and
$\nu _{e} \overline{\nu} _{e} \rightarrow \nu _{e} \overline{\nu} _{e} $ 
the first half of the collision integral is:

\begin{eqnarray}
I_{Coll} & = & \frac{2^{5} G_{F}^{2}}{2E_{1}} \int
\frac{ d^{3}q_{2} d^{3}q_{3} 
d^{3}q_{4}}{(2 \pi)^{9} 8 E_{2} E_{3} E_{4}} (2 \pi)^{4} 
\delta^{4}(q_{1} + q_{2} - q_{3} - q_{4}) 
\cdot F(f) \cdot \nonumber\\
         &   & \cdot \left[ 2 (q_{1} \cdot q_{2})^{2} + 
4  (q_{1} \cdot q_{4})^{2} \right] \nonumber \\
         & = & \frac{20 G_{F}^{2}}{9 \pi ^{3}} E_{1} 
\int dE_{2} \; E_{2}^{2} \cdot F(f),
\nonumber 
\end{eqnarray}
where $F(f) = \left[ - f_{(in \; 1)} f_{(in \; 2)}  \right]$ with 
$ f_{in}$ being the distribution functions for the incoming particles.

For the first 5 reactions in Table 
\ref{table:amplitudes-nu-e} (including neutrinos only) we obtain the result:
\begin{equation}
I_{coll} ^{first \; 5} = \frac{8 G_{F}^{2}}{3 \pi ^{3}} 
\int d p_{2} \;  p_{2}^{3} \left[
- f_{\nu_{e}}(1) \left( f_{\nu_{e}}(2) + \frac{2}{3}  f_{\nu_{\mu}}(2)
\right) \right] .
\end{equation}

The reaction $\nu _{e} \overline{\nu} _{e} \rightarrow e^{+} e^{-}$ 
gives:
\begin{eqnarray}
I_{Coll}^{\nu _{e} \overline{\nu} _{e}} 
&  =  & \frac{G_{F}^{2}}{\pi ^{3}} E_{1} 
\int d \xi d p_{2} \; p_{2}^{3} 
\sqrt{1 - \frac{4 m_{e}^{2}}{s}} (1 - \xi)^2 \cdot \left[ 
- f_{\nu_{e}}(1)  f_{\nu_{e}}(2)\right] \nonumber \\
&     & \cdot \left[ \frac{1}{3} (g_{L}^{2} + g_{R}^{2})
(1 - \frac{m_{e}^{2}}{s})
+2 g_{L}g_{R} \frac{m_{e}^{2}}{s} \right], 
\end{eqnarray}
where $s = 2 E_{1} E_{2} (1 - \xi)$ is expressed trough the angle
$ \xi = \cos \theta _{12}$ between the incoming particles. 

The reactions $\nu_{e} e^{\pm} \rightarrow \nu_{e} e^{\pm} $ give:
\begin{eqnarray}
I_{coll} ^{\nu_{e} e^{\pm}} &=& \frac{G_{F}^{2}}{\pi ^{3}} 
\int d \xi d p_{2} \; p_{2}^{2} \; 
\frac{1 - \frac{m_{e}^{2}}{s}}{E_{1} E_{2}} \; (p_{1} \cdot p_{2})^{2}
\left[ - f_{\nu_{e}}(1)  f_{e}(2) \right] 
\cdot \nonumber \\ 
 & \cdot & 
\left[ \frac{1}{3} (g_{L}^{2} + g_{R}^{2}) 
(4 + \frac{m_{e}^{2}}{s} (1 +  \frac{m_{e}^{2}}{s}) )
- 2 g_{L}g_{R} \frac{m_{e}^{2}}{s}
\right]  
\end{eqnarray}
where $s = 2 E_{1} E_{2} (1 - \beta \xi) + m_{e}^{2}$ and $\beta$ is
the velocity of the incoming electron (positron).
As always we have defined $g_{L} =  1/2 + \sin^{2}\theta_{W}$ and
$g_{R} = \sin^{2} \theta_{W}$.
The angle integrals are easily done, giving for the  $\nu_{e} e^{\pm} $ case:
\begin{eqnarray}
I_{coll}^{\nu_{e} e^{\pm}} 
& = & \frac{G_{F}^{2}}{16 \pi ^{2}} 
\int _{0} ^{\infty} d p_{2} \;  
\frac{p_{2}}{E_{1} ^{2} E_{2}}   \left[ - f_{\nu_{e}}(1)  
f_{e}(2) \right] m_{e}^{6}
\nonumber \\
&   &  \cdot  \left[ 2  (g_{L}^{2} + g_{R}^{2}) \left(
\frac{1}{6 u^2} + \frac{1}{u} + \frac{10 u}{3} - \frac{11 u^2}{6} +
\frac{4 u^3}{9} - \frac{4}{3} \log(u) \right) \right. \nonumber \\
&   & \left. +  4  g_{L}g_{R}  \left( \frac{2}{3 u}  - 3 u 
+ \frac{u^{2}}{2} + 3 \log(u)  \right) \right] ,
\end{eqnarray}
where $u = s/ m_{e}^{2}$.

Similarly  for the interactions of muonic neutrinos with themselves or other
neutrinos, 
which are the first 5 reactions in Table \ref{table:amplitudes-nu-mu}, we get:
\begin{equation}
I_{coll}^{first \;5 \; \nu_{\mu}} =  \frac{8 G_{F}^{2}}{9 \pi ^2} 
E_{1} \int d p_{2} \; p_{2}^3 \left[
- f_{\nu_{\mu}}(1) \left( 4f_{\nu_{\mu}}(2) + f_{\nu_{e}}(2) \right)
\right].
\end{equation}

For the interactions of $\nu_{\mu}$ with electrons we obtain the same 
expression as for  
$\nu_{e}$ except for the change $g_{L} \rightarrow \tilde{g}_{L}
= g_{L} - 1$ and the substitution like
$f_{\nu_{e}} \rightarrow f_{\nu_{\mu}}$.

~
\begin{table}
\begin{center}
\begin{tabular}{cccc|cc} 
& Process &&&& $2^{-5} G^{-2}_{F} S \left| A \right| ^{2}$
\\ \hline\hline
$\nu _{e} + \bar{\nu}_{e} $ & $\rightarrow$ & $  \nu _{e} + \bar{\nu}_{e} $&&&
$4 (p_{1} \cdot p_{4}) (p_{2} \cdot p_{3})$ \\

$\nu _{e} + \nu_{e} $ & $\rightarrow$ & $  \nu _{e} + \nu_{e}$ &&&
$2  (p_{1} \cdot p_{2}) (p_{3} \cdot p_{4})$\\

$\nu _{e}+\bar{\nu}_{e} $ & $\rightarrow$ &
$ \nu _{\mu (\tau)}+\bar{\nu}_{\mu (\tau)} $&&&
$ (p_{1} \cdot p_{4}) (p_{2} \cdot p_{3}) $\\

$\nu_{e} + \bar{\nu}_{\mu (\tau)} $ & $\rightarrow$&
$\nu _{e}+\bar{\nu}_{\mu (\tau)} $&&&
$ (p_{1} \cdot p_{4}) (p_{2} \cdot p_{3})$\\

$\nu _{e} + \nu_{\mu (\tau)} $ & $\rightarrow$ &
$ \nu _{e}+\nu_{\mu (\tau)} $&&&
$ (p_{1} \cdot p_{2}) (p_{3} \cdot p_{4})$\\

$\nu _{e} + \bar{\nu}_{e} $ & $\rightarrow$ & $  e^{+} + e^{-}$ &&&
$4[ ( g_{L}^{2} (p_{1} \cdot p_{4}) (p_{2} \cdot p_{3}) $\\
&&&&&$+ g_{R}^{2}  (p_{1} \cdot p_{3}) (p_{2} \cdot p_{4})$ \\
&&&&&$+ g_{L} g_{R} m _{e}^{2} (p_{1} \cdot p_{2})]$\\

$\nu _{e} +  e^{-} $ & $\rightarrow$ & $ \nu _{e} +  e^{-} $&&&
$4 [ g_{L}^{2} (p_{1} \cdot p_{2}) (p_{3} \cdot p_{4})$ \\
&&&&&$+g_{R}^{2}  (p_{1} \cdot p_{4}) (p_{2} \cdot p_{3})$ \\ 
&&&&&$-g_{L} g_{R} m _{e}^{2} (p_{1} \cdot p_{3}) ] $\\

$\nu _{e} +  e^{+} $ & $\rightarrow$ & $ \nu _{e} +  e^{+} $&&&

$4 [ g_{R} ^{2} (p_{1} \cdot p_{2}) (p_{3} \cdot p_{4})$ \\ 
&&&&&$+g_{L}^{2} (p_{1} \cdot p_{4}) (p_{2} \cdot p_{3}) $ \\
&&&&&$-g_{L} g_{R} m _{e}^{2} (p_{1} \cdot p_{3}) ] $
\end{tabular}
\caption{The matrix elements for various processes with electronic neutrinos;
here $g_{L} =  1/2 + \sin^{2}\theta_{W}$ and
$g_{R} = \sin^{2} \theta_{W}$.}
\label{table:amplitudes-nu-e}
\end{center}

\begin{center}
\begin{tabular}{cccc|cc} 
& Process &&&& $2^{-5} G^{-2}_{F} S \left| A \right| ^{2}$
\\ \hline\hline
$\nu _{\mu} + \bar{\nu}_{\mu} $ & $\rightarrow$ & $  \nu _{\mu} + \bar{\nu}_{\mu} $&&&
$4 (p_{1} \cdot p_{4}) (p_{2} \cdot p_{3})$ \\

$\nu _{\mu} + \nu_{\mu} $ & $\rightarrow$ & $  \nu _{\mu} + \nu_{\mu}$ &&&
$2  (p_{1} \cdot p_{2}) (p_{3} \cdot p_{4})$\\

$\nu _{\mu}+\bar{\nu}_{\mu} $ & $\rightarrow$ &
$ \nu _{e (\tau)}+\bar{\nu}_{e (\tau)} $&&&
$ (p_{1} \cdot p_{4}) (p_{2} \cdot p_{3}) $\\

$\nu_{\mu} + \bar{\nu}_{e (\tau)} $ & $\rightarrow$&
$\nu _{\mu}+\bar{\nu}_{e (\tau)} $&&&
$ (p_{1} \cdot p_{4}) (p_{2} \cdot p_{3})$\\

$\nu _{\mu} + \nu_{e (\tau)} $ & $\rightarrow$ &
$ \nu _{\mu}+\nu_{e (\tau)} $&&&
$ (p_{1} \cdot p_{2}) (p_{3} \cdot p_{4})$\\

$\nu _{\mu} + \bar{\nu}_{\mu} $ & $\rightarrow$ & $  e^{+} + e^{-}$ &&&
$4[ ( \tilde{g}_{L}^{2} (p_{1} \cdot p_{4}) (p_{2} \cdot p_{3}) $\\
&&&&&$+ g_{R}^{2}  (p_{1} \cdot p_{3}) (p_{2} \cdot p_{4})$ \\
&&&&&$+ \tilde{g}_{L} g_{R} m _{e}^{2} (p_{1} \cdot p_{2})]$\\

$\nu _{\mu} +  e^{-} $ & $\rightarrow$ & $ \nu _{\mu} +  e^{-} $&&&
$4 [ \tilde{g}_{L}^{2} (p_{1} \cdot p_{2}) (p_{3} \cdot p_{4})$ \\
&&&&&$+g_{R}^{2}  (p_{1} \cdot p_{4}) (p_{2} \cdot p_{3})$ \\ 
&&&&&$-\tilde{g}_{L} g_{R} m _{e}^{2} (p_{1} \cdot p_{3}) ] $\\

$\nu _{\mu} +  e^{+} $ & $\rightarrow$ & $ \nu _{\mu} +  e^{+} $&&&

$4 [ g_{R} ^{2} (p_{1} \cdot p_{2}) (p_{3} \cdot p_{4})$ \\ 
&&&&&$+\tilde{g}_{L}^{2} (p_{1} \cdot p_{4}) (p_{2} \cdot p_{3}) $ \\
&&&&&$-\tilde{g}_{L} g_{R} m _{e}^{2} (p_{1} \cdot p_{3}) ] $
\end{tabular}
\caption{The matrix elements for various processes with muonic or tau
neutrinos; here $\tilde{g}_{L} =g_{L} - 1 = 
- 1/2  + \sin^{2}\theta_{W}$ and
$g_{R} = \sin^{2} \theta_{W}$.}
\label{table:amplitudes-nu-mu}
\end{center}
\end{table}

\newpage
\begin{table}
\begin{center}
\begin{tabular}{|c||c|c|c|}
\hline
 
&&&\\
Program&
~~$T_\gamma/T_\nu$~~&~~$\delta\rho_{\nu_e}/\rho_{\nu_e}
 $~~&~~$\delta\rho_{\nu_{\mu(\tau)}}/\rho_{\nu_{\mu(\tau)}} $~~\\
&&&\\
\hline
\hline
Full distribution f(x,y)&&&\\
momentum points 100&&&\\
initial time $x_{in}=0.2$& 1.3996& 0.80 \%  & 0.26 \% \\
Bulirsch-Stoer method \cite{numrec} &&&\\
\hline
Full distribution f(x,y)&&&\\
momentum points 100&&&\\
initial time $x_{in}=0.1$& 1.3996& 0.79 \%  & 0.25 \% \\
Simple time evolution&&&\\
\hline
Full distribution f(x,y)&&&\\
momentum points 200&&&\\
initial time $x_{in}=0.1$& 1.3994& 0.83 \%  & 0.33 \% \\
Simple time evolution&&&\\
\hline
Correction $\delta(x,y)$&&&\\
momentum points 200&&&\\
initial time $x_{in}=0.1$& 1.3991& 0.94 \%  & 0.40 \% \\
Simple time evolution&&&\\
\hline
Same, but assuming&&&\\
entropy conservation&&&\\
(photon cooling due to& 1.4010& 1.13 \%  & 0.53 \% \\
$\nu e$-interaction is neglected )&&&\\
\hline
\hline
\end{tabular}
\bigskip
\caption{Comparison of the results of different ways of calculation. All the 
results are for FD case.}
 \end{center}
\end{table}

\bigskip

\clearpage

\newpage
{\large \bf Figure Captions:}
\vskip1cm
\noindent
{\bf Fig. 1} $~~~$ The ratio $T_{\gamma}/T_{\nu}$ as  
a function of the dimensionless "time"
$x=m a$  for the cases of Fermi-Dirac and Maxwell-Boltzmann statistics. 
The value $x=1$ corresponds to $T_{\nu}=1 MeV$. 

\vskip1cm
\noindent
{\bf Fig. 2} $~~~$ The relative distortion of the total neutrino energy density
$\delta \rho / \rho_{eq}$ (for the definition see eq. (\ref{drhonu}))
 as a function of 
$x=m a$ for the case of 
"entropy conservation" (upper curve) and 
energy conservation according to eq. (\ref{drhodx}) (lower curve). 
In the first case we 
neglect the energy exchange between neutrinos and $e^\pm$ which is 
equivalent to the requirement of  entropy conservation in the 
electromagnetic plasma.

\vskip1cm
\noindent
{\bf Fig. 3} $~~~$ The "time"  evolution of the correction to the neutrino
distribution functions $\delta_j =  
(f_{\nu_j} -f_{\nu}^{eq})/{f_{\nu}^{eq}}$ 
for several fixed values of momentum $y=3,5,7$. 
In Fig.3a and 3b  we present respectively the 
results for electronic neutrinos, and for muonic (tau) ones.

\vskip1cm
\noindent
{\bf Fig. 4} $~~~$ The evolution of the non-equilibrium corrections to the
distribution functions for the momentum value $y=5$  for the 
electronic (solid curve) and muonic (tau) (dotted curve) neutrinos 
in the cases of  FD and MB statistics. 
 
\vskip1cm
\noindent
{\bf Fig. 5} $~~~$ The distortion of the neutrino spectra
$\delta_j = (f_{\nu_j}-f_\nu^{eq})/f_\nu^{eq} $
as functions of the dimensionless momentum $y$ at the final "time" $x=60$.
The dashed lines $a$ and $c$ correspond to Maxwell-Boltzmann statistics,
while the solid lines $b$ and $d$ correspond to Fermi-Dirac statistics.
The upper curves $a$ and $b$ are for electronic neutrinos, while the lower 
curves $c$ and $d$ are for muonic (tau) neutrinos. All the curves can be 
well approximated by a second order polynomial 
in $y$ like $\delta = A y (y -B)$.

\newpage
\psfig{file=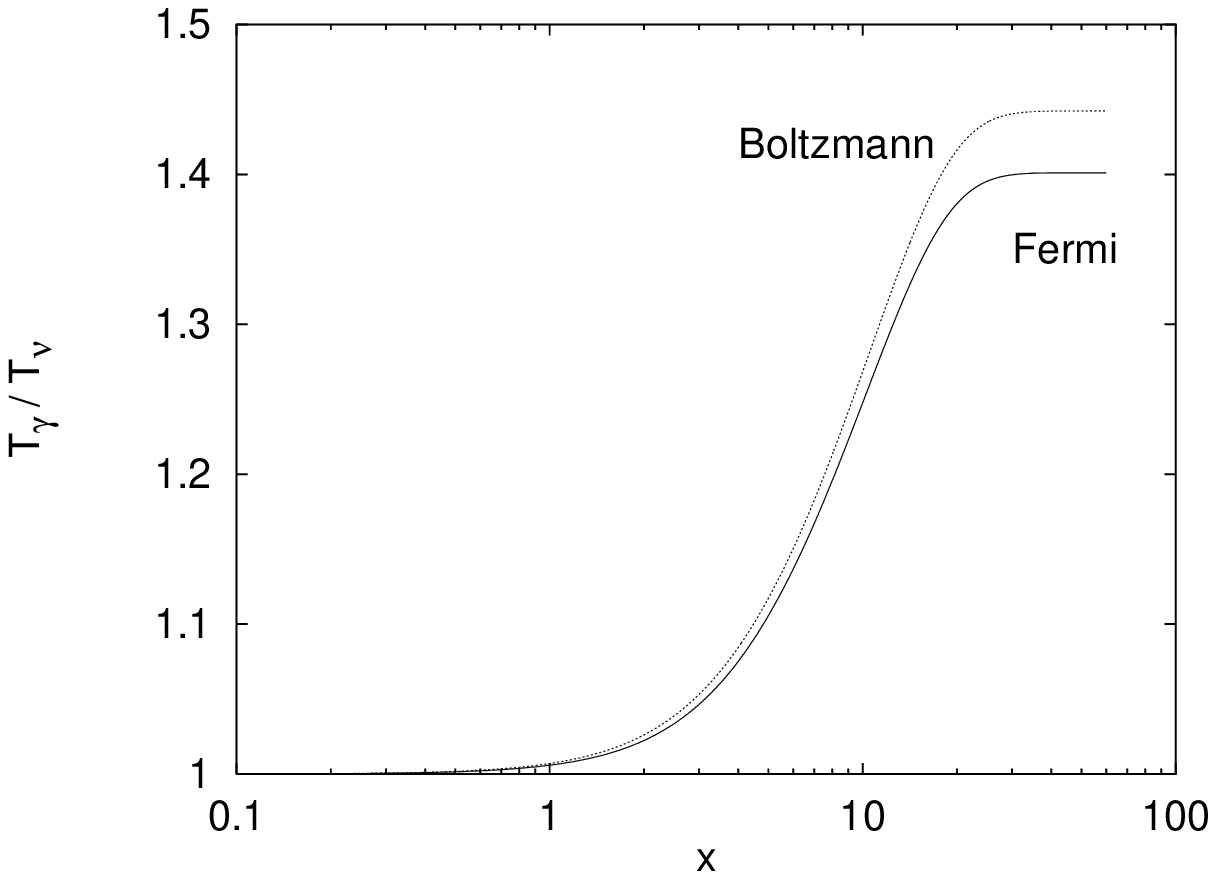,width=5in,height=3.5in}
\begin{center}
{\bf Figure 1.}
\end{center}

\psfig{file=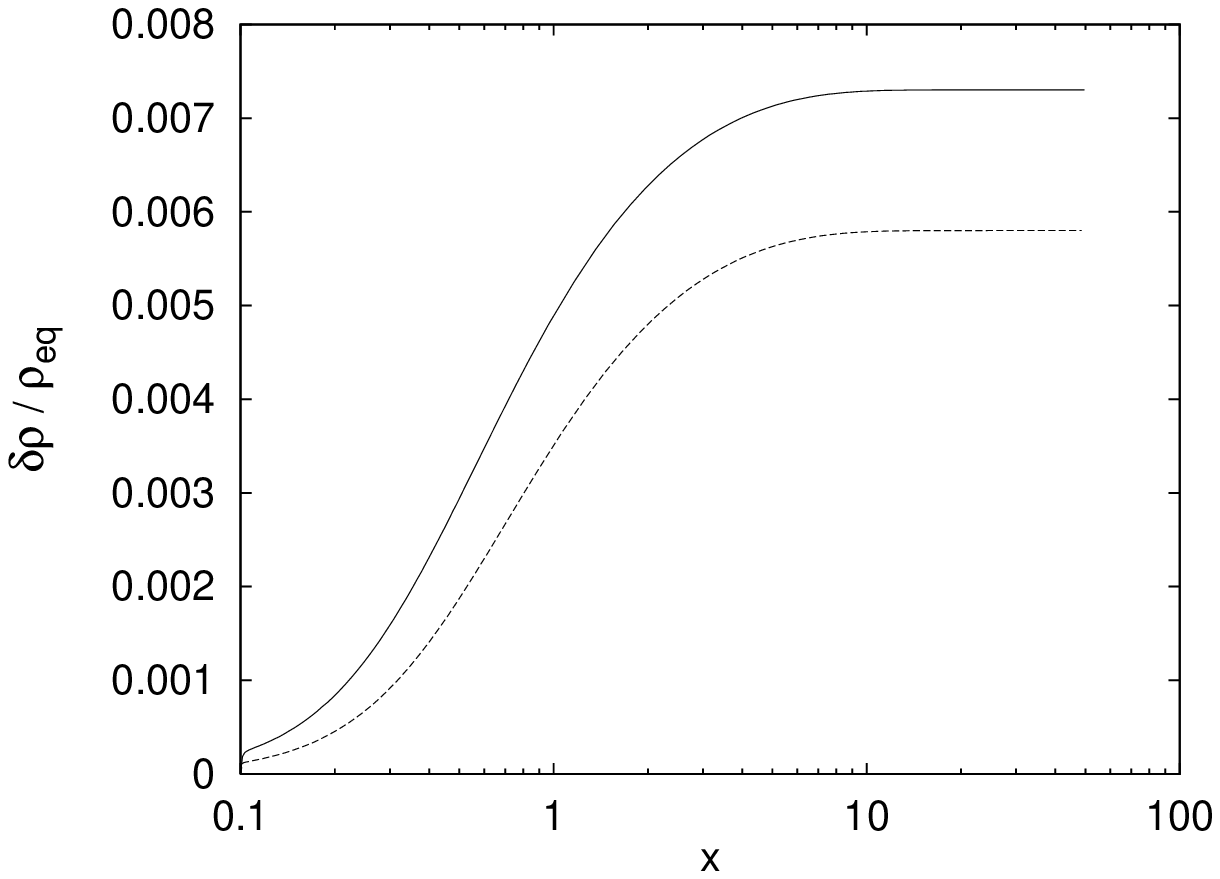,width=5in,height=3.5in}
\begin{center}
{\bf Figure 2.}
\end{center}

\newpage
\psfig{file=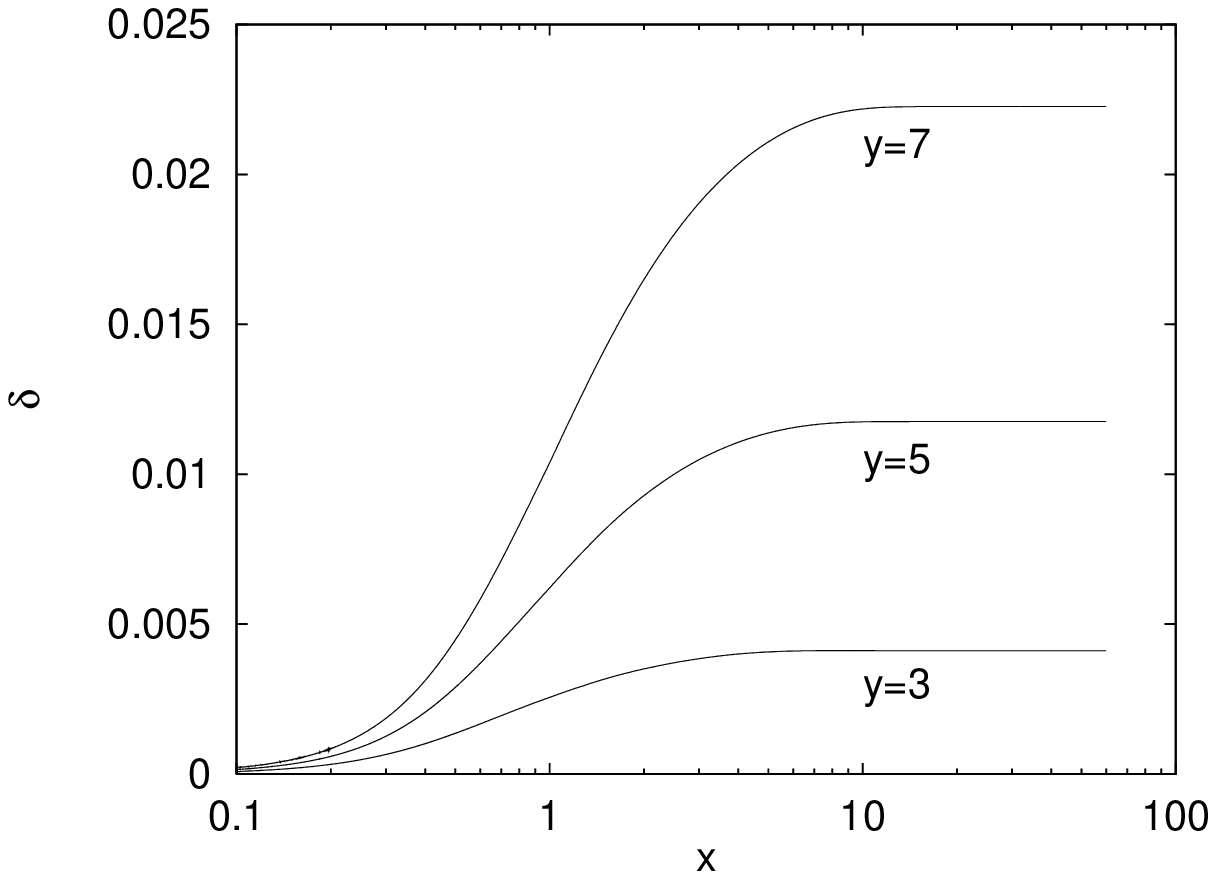,width=5in,height=3.5in}
\begin{center}
{\bf Figure 3a.}
\end{center}

\psfig{file=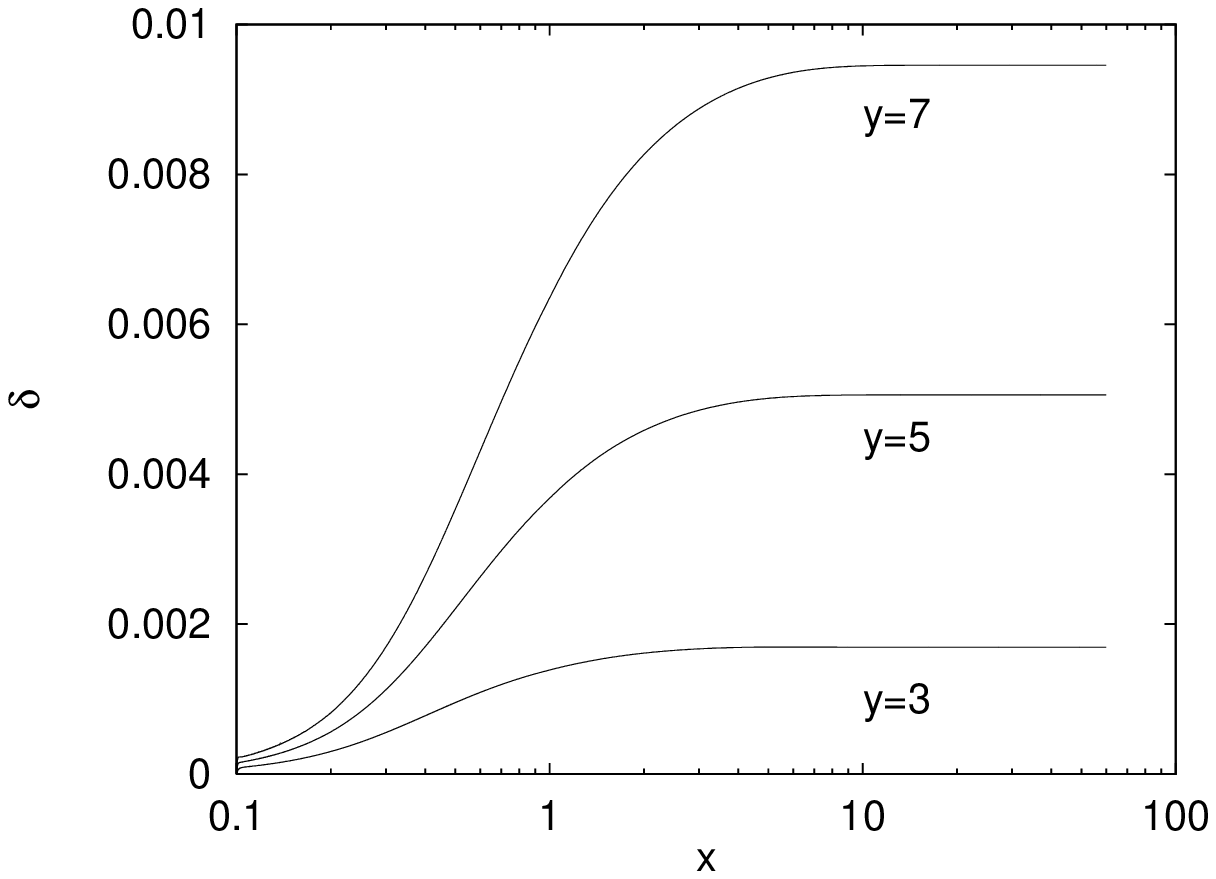,width=5in,height=3.5in}
\begin{center}
{\bf Figure 3b.}
\end{center}

\newpage
\psfig{file=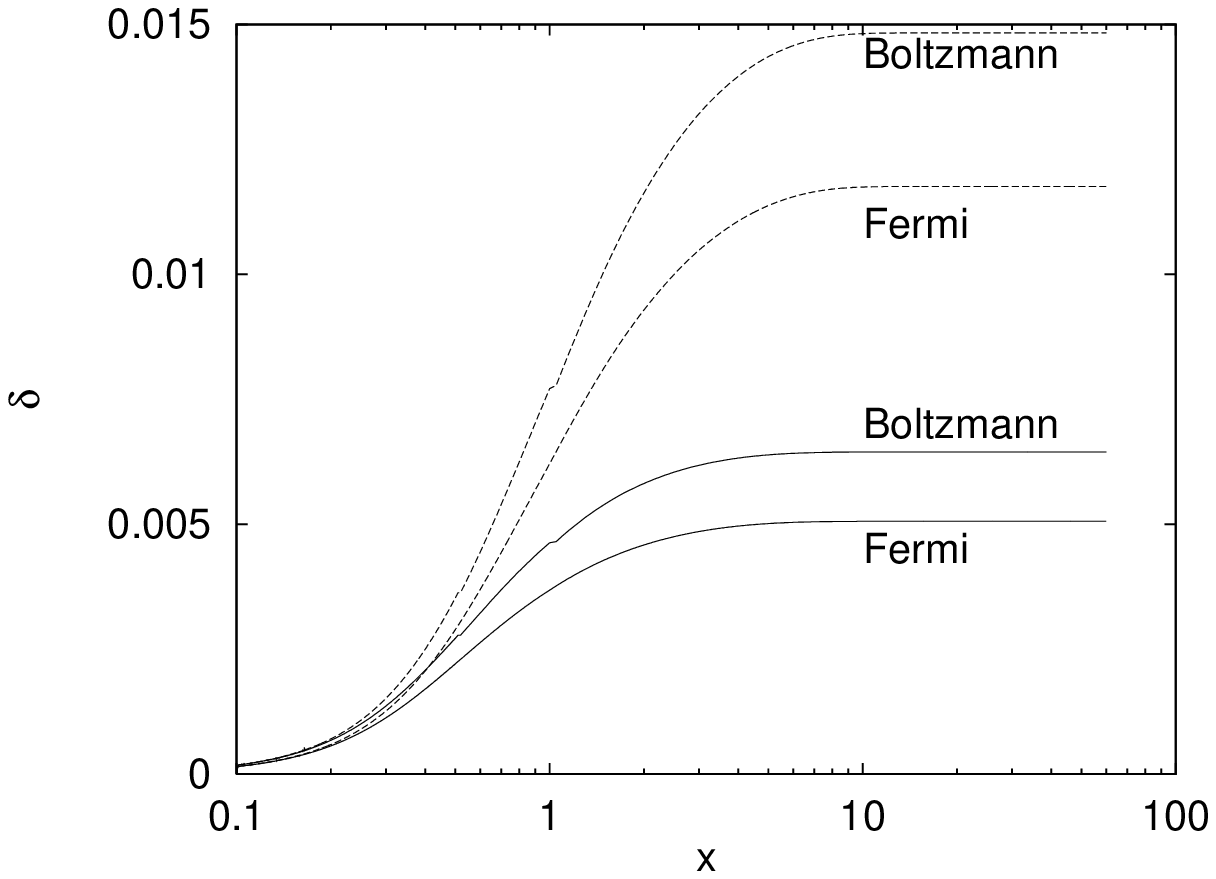,width=5in,height=3.5in}
\begin{center}
{\bf Figure 4.}
\end{center}

\psfig{file=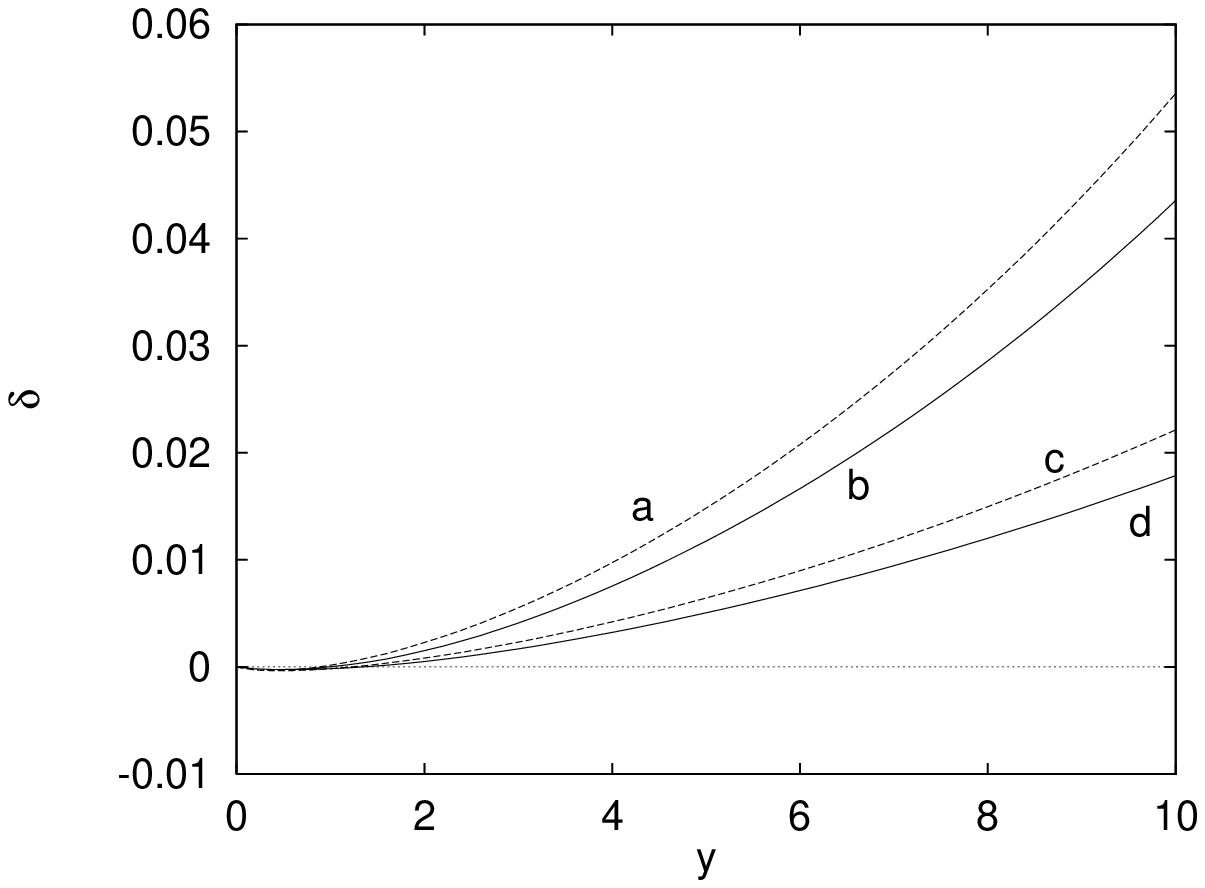,width=5in,height=3.5in}
\begin{center}
{\bf Figure 5.}
\end{center}

\end{document}